%% file: main.tex
\documentclass[sigconf]{acmart}

\usepackage[hide]{chato-notes}

\setcopyright{acmcopyright}
\copyrightyear{2020}
\acmYear{2020}
\acmDOI{10.1145/xxxxxxx.xxxxxxx}

\acmConference[CIKM '20]{CIKM '20: ACM Conference on Information and Knowledge Managenent}{October 19--23, 2020}{Galway, Ireland}
\acmBooktitle{CIKM '20: ACM Conference on Information and Knowledge Management, October 19--23, 2020, Galway, Ireland}
\acmPrice{15.00}
\acmISBN{978-1-4503-XXXX-X/18/06}

\usepackage{graphicx}  
\usepackage{paralist}  
\usepackage{booktabs}  
\usepackage{algorithm}   
\usepackage{algorithmic} 
\usepackage{xcolor}      
\usepackage{url}         
\usepackage{amsmath}	
\usepackage{xspace}   
\usepackage{subfig}   

\input{macros}

\newtheorem{problem}{Problem}

\begin{document}


\newcommand{\runningtitle}{Intersectional Affirmative Action Policies for Top-k Candidates Selection}
\newcommand{\maintitle}{Intersectional Affirmative Action Policies for Top-k~Candidates~Selection}
\title[\runningtitle]{\maintitle}

%

\author{Michael Mathioudakis}
\affiliation{%
  \institution{University of Helsinki}
}
\author{Carlos Castillo}
\affiliation{%
  \institution{Universitat Pompeu Fabra}
}
\author{Giorgio Barnabo}
\affiliation{%
  \institution{Sapienza University of Rome}
}
\author{Sergio Celis}
\affiliation{%
  \institution{Universidad de Chile}
}

\renewcommand{\shortauthors}{Mathioudakis, Castillo, Barnabo, Celis}




\input{00-abstract}


\maketitle              


\input{01-introduction}
\input{02-related-work}
\input{03-problem-statement}
\input{04-algorithms}
\input{05-experiments}
\input{06-results}
\input{07-conclusions}
\input{08-acknowledgements}

\balance
\clearpage

\bibliographystyle{splncs04}
\bibliography{biblio.bib}

\end{document}

%% file: macros.tex
\usepackage[acronym]{glossaries}

\newcommand{\hide}[1]{}

\newcommand{\spara}[1]{\smallskip\noindent{\bf{#1}}}

\newcommand{\budget}{\ensuremath{k}\xspace}
\newcommand{\auxtable}{\ensuremath{R}\xspace}

\newcommand{\napplicants}{\ensuremath{n}\xspace}
\newcommand{\utility}{\ensuremath{u}\xspace}
\newcommand{\setutility}{\ensuremath{U}\xspace}
\newcommand{\selectionrate}{\ensuremath{p}\xspace}
\newcommand{\selected}{\ensuremath{x}\xspace} 
\newcommand{\setselected}{\ensuremath{X}\xspace}
\newcommand{\classes}{\ensuremath{C}\xspace}
\newcommand{\nclasses}{\ensuremath{|C|}\xspace}
\newcommand{\objective}{\ensuremath{\mathbf{J}}\xspace}
\newcommand{\weight}{\ensuremath{\lambda}}
\newcommand{\totalutility}{\ensuremath{B}\xspace}
\newcommand{\totaldiscrepancy}{\ensuremath{D}\xspace}

\newcommand{\singletrack}{\texttt{Single-track}\xspace}
\newcommand{\separatetrack}{\texttt{Separate-tracks}\xspace}

\newacronym{dp}{{\sc DynPr}}{{Dynamic programming algorithm}}
\newacronym{greedy}{{\sc Greedy}}{{Greedy algorithm}}

%% file: 00-abstract.tex

\begin{abstract}
We study the problem of selecting the top-k candidates from a pool of applicants, where each candidate is associated with a score indicating his/her aptitude.
Depending on the specific scenario, such as job search or college admissions, these scores may be the results of standardized tests or other predictors of future performance and utility.
We consider a situation in which some groups of candidates experience historical and present disadvantage that makes their chances of being accepted much lower than other groups. In these circumstances, we wish to apply an affirmative action policy to reduce acceptance rate disparities, while avoiding any large decrease in the aptitude of the candidates that are eventually selected.
Our algorithmic design is motivated by the frequently observed phenomenon that discrimination disproportionately affects individuals who simultaneously belong to multiple disadvantaged groups, defined along intersecting dimensions such as gender, race, sexual orientation, socio-economic status, and disability.
%
%
In short, our algorithm's objective is to simultaneously
\begin{inparaenum}[(i)]
\item select candidates with high utility, and
\item level up the representation of disadvantaged intersectional classes.
\end{inparaenum}
This naturally involves trade-offs and is computationally challenging due to the the combinatorial explosion of potential subgroups as more attributes are considered.
We propose two algorithms to solve this problem, analyze them, and evaluate them experimentally using a dataset of university application scores and admissions to bachelor degrees in an OECD country.
%
%
Our conclusion is that it is possible to significantly reduce disparities in admission rates affecting intersectional classes with a small loss in terms of selected candidate aptitude.
To the best of our knowledge, we are the first to study fairness constraints with regards to intersectional classes in the context of top-k selection.
\end{abstract}

%% file: 01-introduction.tex

\section{Introduction}
\label{sec:introduction}

Avoiding algorithmic discrimination in contemporary information systems has become a major issue in all areas of computing research~\cite{barocas2016big}.
Data mining, machine learning, artificial intelligence, and information retrieval researchers - often in collaboration with law scholars and social scientists - have been developing principles, methods and techniques to detect and mitigate discriminatory algorithmic biases~\cite{hajian2016algorithmic}.
Indeed, with the increasing automation of decision processes in all aspects of human life, avoiding unfair and unacceptable disadvantage for specific individuals and groups has turned into an urgent political objective, as well as a technical challenge.
Recently, intersectionality theory~\cite{crenshaw1989demarginalizing,foulds2018intersectional,cabrera2019fairvis} has enriched the debate on algorithmic fairness by showing how often discrimination affects people who lay at the intersection of several protected attributes. This finding should in turn lead to more effective actions in this emerging research field.

\subsection{Intersectional Effects in University Admissions}
To practically show how several protected attributes can combine into strongly underprivileged or privileged intersectional classes, let us look at an example from our dataset, which is described in detail in Section~\ref{sec:experiments}.
This dataset represents the centralized admission process to universities that took place in an OECD country in 2017.\footnote{Country name withheld for double-blind review.}
It contains anonymized data with demographic, socio-economic and educational information of about 300,000 high school graduates who were eligible to pursuit a bachelor degree in a university.
Only 120,000 of them took the required admission tests, and about 80,000 of them were admitted to a program.

The regular admission process for undergraduate programs in this country is based on a linear combination of standardized test scores and high school grades.
Each program weights these quantities differently. 
Students apply by listing up to ten programs in decreasing order of preference.
Each program offers a number of vacancies and admits the applicants with the highest scores to fill those vacancies.
Once a student is accepted at one of his/her preferred programs, the next items in his/her preference list are ignored.
In general, admission scores go from 0 to 850 points. Most successful applicants have scores above 450 points, but usually 750 points or more are needed to be accepted into the most prestigious and competitive programs such as medicine, law, or engineering.

To highlight the complex discrimination mosaics behind this admission system, we selected the following three protected attributes, coded according to Table~\ref{tbl:coding}.

\begin{table}[h]
\caption{Coding of inter-sectional groups.}
\label{tbl:coding}
\begin{tabular}{ll}\toprule
Socio-economic status & \textbf{1.} High income \\
  & \textbf{2.} Medium income \\
  & \textbf{3.} Low income \\
  \midrule
High-school type & \textbf{A.} Private \\
 & \textbf{B.} Public or subsidized \\
  \midrule
Regional development & \textbf{a.} High-development region \\
 & \textbf{b.} Low-development region \\
 \bottomrule
\end{tabular}
\end{table}

\begin{itemize}
\item \textbf{Socio-Economic Status:} based on the income decile of the applicant's family, we split students into 
high income (1), 1th to 3rd decile,
medium income (2), 4th to 6th decile,
and
low income (3), 7th to 10th decile;
\item \textbf{High School Type:} students are divided into those who went to a
\emph{private} (A) high school and those who went to a
\emph{public or subsidized} (B) one;
\item \textbf{Regional Development: } according to 2017 Human Development Index data for this country,\footnote{Reference omitted for double-blind review.} we divide students into those coming from \emph{high-development areas/regions} (a), which includes most large metropolitan areas, and those from \emph{low-development areas/regions} (b). 
\end{itemize}

The intersection of these attributes gives rise to 12 intersectional classes. The most privileged group is coded as ``1Aa'' (high income, private high school, high-development region), and the most disadvantaged group as ``3Bb'' (low income, public/subsidized high school, low-development region).

\begin{figure*}[h]
\centering\includegraphics[width=0.95\columnwidth]{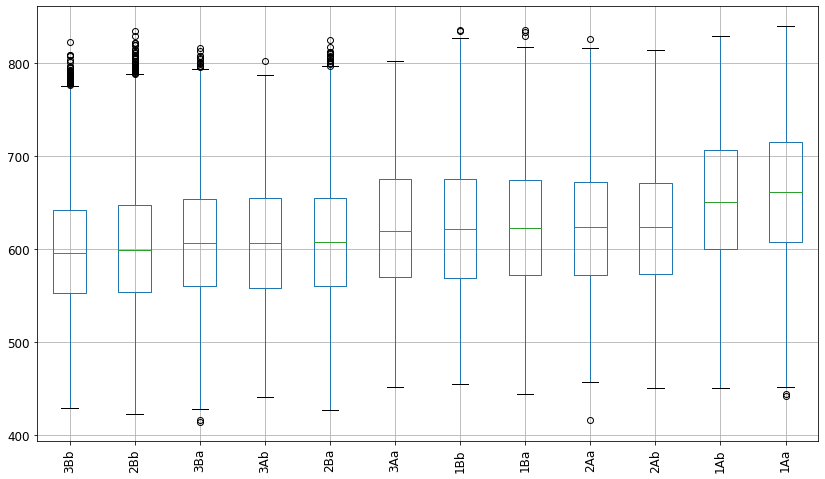}
\caption{Range of scores obtained by groups at the intersection of socio-economic status, high school type, and regional development.}
\label{fig:motivating-example}
\end{figure*}

The plot of Figure~\ref{fig:motivating-example},
which considers only the applicants who were eventually admitted to some university program,
shows the distribution of admission scores among students of each class.
As we can see, the twelve intersectional classes have substantially different admission score distributions.
For instance, the average score decreases by more than 60 points from
the most privileged group (1Aa - High Income, Private High School, High-Development Region) to
the most underprivileged group (3Bb - Low Income, Public or Subsidized High School, Low-Development Region).
It is important to note that often just a few points can determine a student's chances to get admitted to a specific program. 

By taking a closer look, we can then a complex intersectional structure of privilege, for example:

\begin{itemize}
\item coming from a low-development region does not make much of a difference for students from families with high income, while is has a moderate effect--10 points on average--on low-income students;
\item going to a private school gives a large advantage (+35 points) to high-income students living in a highly developed region, but not as much (+10 points) to low-income students;
\item in high-development regions, in public or subsidized high schools, having low income gives a moderate disadvantage (-16 points), that worsens if you live in a low-development region (-25 points);
\item both in low- and high- development regions, going to a private high school gives a large advantage only if you come from a high-income family (+38 or +43 points of difference);
\end{itemize}

In a nutshell, what we observe is that two opposite forms of discrimination take place.
On the one hand, the compound effect of having a low income in a low-development region and going to a public or subsidized high school makes this specific intersectional class the most underprivileged and shows that these three disadvantages reinforce each other. 
On the other hand, having a high income, going to a private high school and living in a high-development region generates a strong condition of privilege and is not just the sum of these advantages.

For instance, all other things being equal, suppose having low income decreases an applicant's expected score by 10 points with respect to his or her high income counterparts, while going to a public high school induces a 5 point drop in average outcome with respect to private high school students.
In an intersectional scenario, on average more than 15 points of difference would be expected between poor students going to a public high school and rich students going to private high schools. Such effects may well reduce social mobility and facilitate the perpetuation of privilege.

\subsection{Affirmative Action in Top-K Selection}

The algorithmic problem of considering group representation criteria when selecting the top k individuals from a pool of candidates for an educational program, a job position, or a scholarship has been widely studied~\cite{zehlike2017fa, sapiezynski2019quantifying, geyik2019fairness, gorwa2020algorithmic}.
This choice is generally made on the grounds of some future performance indicators or standardized tests. Given a predictor trained on historical data, a ranking is generated and only people in the top k slots are selected.
Frequently, though, these selection processes give rise to subtle forms of discrimination, mainly due to historical legacies that block some social groups from performing well in the relevant indicators and tests.
For this reason, when designing such systems, it is often recommended to design appropriate constraints that would eventually adjust disparities, particularly in prestigious institutions from which underprivileged students are more often marginalized \cite{hinrichs2012effects}.

From a technical standpoint, one of the most widespread ways of measuring algorithmic fairness is through the notion of \emph{demographic parity} or \emph{statistical parity} \cite{kamiran2009classifying}.
This notion implies that the probability of a candidate to be selected is \emph{independent} from the whole set of protected attributes. In other words, we want that--on average--people from all intersectional classes have the same chance of getting admitted.
These corrective interventions usually go under the name of \emph{affirmative action} or \emph{positive action} policies. More specifically, in the case of top-k selection policies, there are two essential objectives hoped to be achieved through the implementation of an affirmative action:
\begin{inparaenum}[(i)]
  \item to select candidates who have a high expected performance, and
  \item to ensure candidates from disadvantaged backgrounds are well represented.
\end{inparaenum}
Both properties meet some notion of fairness: the former in favor of general utility, the latter against discrimination not related to performance.

As can be easily guessed, however, an intricate and deeply rooted historic system of discrimination makes the actual admission process disadvantageous for some intersectional social groups.
In other words, a student's position in the final ranking and his/her chances of being admitted to a university program are correlated with cultural, social and economic protected attributes that should not play a role in the process.
Consequently, alongside the regular admission process, many countries and universities put in place additional affirmative action programs intended for students who come from disadvantaged backgrounds \cite{posselt2012access, black2020winners}.

\subsection{Our contribution}

In this paper, we formalize the problem of mitigating intersectional discrimination in admission policies and present algorithms to address it.
These algorithms maximize the total utility for the system, expressed as the sum of all the admission scores of the selected students, while reducing the discrepancies between the admission rate of underprivileged and privileged intersectional classes.
We demonstrate the effectiveness of these methods by performing several experiments in the context of university admissions.

The rest of this paper is organized as follows.
Section~\ref{sec:related} overviews previous work related to ours.
Section~\ref{sec:statement} introduces formally our problem statement.
Section~\ref{sec:algorithms} describes the algorithms we propose, which are experimentally evaluated in Section~\ref{sec:experiments}.
The last section presents our conclusions and suggests future work.

%% file: 02-related-work.tex

\section{Related Work and Background}\label{sec:related}

In this section we summarize previous work on fairness-aware top-k selection and data-driven affirmative action. We also provide some background on intersectionality theory.

\subsection{Fairness-Aware Top-K Selection}

Algorithmic fairness in data mining and machine learning applications is an emerging research area - for a survey, see~\cite{mitchell2018prediction} - and scholars have recently started studying fairness in ranking or top-k selection.
For instance, Celis et al.~\cite{celis2017ranking} and Stoyanovich et al.~\cite{stoyanovich2018online} incorporated fairness constraints in batch processing and online processing respectively. In these works, every element is associated a certain utility and belongs to a specific group. In forming the ranking, constraints on the minimum and maximum number of candidates from every group have to be met. The goal is to rank candidates by decreasing utility without violating the constraints.

In an attempt to determine whether a specific ranking is statistically compatible with a random process that draws candidates from protected and non-protected populations, Zehlike et al.~\cite{zehlike2017fa} propose a fair representation condition based on a binomial test.
Other proposed tests for fair rankings tend to be based on ensuring that members of a protected group are sufficiently included among the top positions in the ranking, i.e., they use some form of positional discount when computing the benefit or visibility that a ranking apportions to members of different groups~\cite{singh2018fairness,geyik2019fairness}.
Current methods to create fair rankings tend to be based on Learning To Rank (LTR), extending the traditional objective in LTR of maximizing ranking similarity with training data with an additional term penalizing disparities in ranking across groups~\cite{singh2019policy,zehlike2020reducing}.

Kearns et al.~\cite{kearns2017meritocratic} studied the case of selecting k candidates drawn from different groups in which candidates across groups are not comparable. Hence, what matters is the relative position of candidates within their groups.
Finally, Mathioudakis et al.~\cite{mathioudakis2019affirmative} used counterfactual analysis to design affirmative action policies that are applied either at the stage of candidate scoring (i.e., giving bonus points to underrepresented groups) or at the stage of decision (i.e., imposing quotas for the acceptance of underrepresented groups).  

In comparison with previous work, we design affirmative action policies for the intersection of multiple protected attributes, and our objective is to reduce disparities in admission rates across intersectional groups.

\subsection{Data-Driven Affirmative Action}

\spara{Downstream effects of affirmative action.} Much research has been conducted on downstream or ``cascade'' effects of affirmative action policies.
From a theoretical standpoint, Hu and Chen~\cite{hu2018short} consider a labor market of either temporary or permanent employers. Temporary employers are required to hire from disadvantaged groups, while permanent employers can hire based on expected utility alone.
The intention is to reduce discrimination by absorbing workers from disadvantaged groups in the temporary market.
Along the same lines, Kannan et al.~\cite{kannan2019downstream} focused their work on affirmative action policies in education, looking at the employability of graduates from disadvantaged groups.
Their goal is to achieve employment rate parity, possibly by creating an incentive for employers to adopt a color-blind hiring policy.
Finally, Mouzannar et al.~\cite{mouzannar2019fair} analyzed the effect of affirmative action policies on the qualifications of different groups in society.

\spara{Data-driven analysis.} It is now well documented that in the last sixty years major universities - both public and private - have been slowly adopting admission mechanisms that unwillingly discriminate historically disadvantaged and underrepresented groups~\cite{posselt2012access, black2020winners}.
Students who come from a privileged background - wealthier, more educated, better integrated into the political and social system - tend to outperform other students even though there are no significant differences in their QI or expected academic performance~\cite{bastedo2017improving}.
In other words, selection processes tend to reproduce systems of power within society. For example, as Wightman showed~\cite{wightman1998lsac} with a large experiment, in the US law school applicants of color had significantly poorer results on admission tests, but once admitted their performance was comparable with that of white students.
To combat this distortion, several affirmative action policies have been put in place worldwide.
Brazil, for example, has promoted bonus and racial quotas~\cite{estevan2019redistribution, francis2012using} that resulted in an increase of both black students and students coming from low socioeconomic backgrounds in major public universities.
Other countries, 
instead, used coefficient-based policies to counterbalance differences in standardized test results between men and women~\cite{arias2016brecha}, as well as between students from private and public high schools~\cite{cornejo2006experimento}, with some success~\cite{larroucau2013efecto}.
In the US, university admission officers tended to admit disadvantaged students at a higher rate when provided with information about a candidate's economic and family background~\cite{bastedo2018we}.

In these and other countries there is abundant data about university admissions, more applicants than vacancies at top institutions, and a need to optimize resources.
This means data-driven policy design can be undertaken to address this need, which is what we describe in this paper through an admission algorithm that is aware of intersectional classes.

\subsection{Background: Intersectionality}

The term \emph{intersectionality} first appeared in a 1989 paper~\cite{crenshaw1989demarginalizing} by civil rights advocate Kimberl{\'e} Crenshaw, who used it to describe those forms of social inequality that stem from interlocking social institutions.
Noticing that the first two waves of feminism mainly focused on middle class white women's struggles, Crenshaw proposed instead to look at how one's multiple identities (e.g., gender, race, class, sexuality, ability) combine to create different and specific modes of discrimination.
It was not by change that in its political agenda~\cite{wood2009gendered} the third wave feminism - which emerged shortly after Crenshaw's paper - paid greater attention to people who are subjected to multiple forms of inequality
\cite{mccall2008complexity}.

More specifically, these new feminists opened up to the idea of addressing black women's specific condition, instead of treating all women as if they experienced the same difficulties.
%
Needless to say, intersectionality has raised criticism due to its difficult practical implementation and the possible risk of paradoxically weakening civil and social rights movements.
As Rebecca Reilly-Cooper and Rekia Jibrin point out, excessive fragmentation and individualization of discriminatory conditions can in fact make very difficult to synthesize a common actionable and to reach legal praxis~\cite{jibrin2015revisiting}
%
as the process towards recognition of previous denied rights involves the mobilization of a large mass of people.

Nevertheless, both the UK and the EU~\cite{lawson2016european} have non-discrimination law that tries to take into account complex and deep rooted forms of intersectional unfair treatment.
The UK, in particular, lists ``age, disability, gender reassignment, marriage or civil partnership, pregnancy and maternity, race, religion or belief, sex and sexual orientation'' as ``protected characteristics'' under the 2010 Equality Act.
Section 14 of this bill contains a provision to cover discrimination on up to two combined grounds. This section, however, never became effective because the government considered it ``too complicated and burdensome for businesses''~\cite{Lawyer}.

With regards to computer science research, just recently scholars started incorporating the concept of intersectionality in their work on algorithmic fairness~\cite{foulds2018intersectional,cabrera2019fairvis}.
Intersectional discrimination has been investigated in the context of
automated facial analysis~\cite{buolamwini2018gender},
expectation constraints~\cite{fitzsimons2018intersectionality},
classification problems~\cite{morina2019auditing}, and
many other fields of artificial intelligence~\cite{hoffmann2019fairness}.

%
In contrast to previous work, this paper develops a top-k candidate selection algorithm specifically designed to mitigate intersectional discrimination.

%% file: 03-problem-statement.tex

\section{Problem Statement}
\label{sec:statement}

In this section we present our notation and formally describe our problem statement.

\subsection{Setting}
\label{sec:setting}

\begin{table}[t]
\caption{Notation}
\label{tbl:notation}
\centering\begin{tabular}{cl}
\toprule
$\napplicants$ & Number of candidates \\
$\selectionrate$ & Selection rate: fraction of candidates to be admitted\\
$\classes$ & Intersectional classes \\
$\napplicants_i$ & Cardinality of the $i^{\text{th}}$ intersectional class \\
$\utility_{ij}$ & Utility of candidate $j$ from intersectional class $i$ \\
$\selected_{ij}$ & Decision for candidate $j$ from class $i$; selected (1) or not (0) \\
$\setutility$ & Set of utilities of all candidates $\setutility = \bigcup\{\utility_{ij}\}$ \\
$\setselected$ & Set of decisions for all candidates $\setselected = \bigcup\{\selected_{ij}\}$ \\
\bottomrule
\end{tabular}
\end{table}

In our setting, a set of candidates are considered for selection by one institution, under the constraint that only the top $k$ of them are to be selected.
This setting arises in many real-life scenarios -- for example, when multiple professionals apply for a public sector job or multiple students apply for a position at a university program.

\spara{Intersectional classes.}
Each individual is assumed to be associated with certain protected attributes, such as gender, race, sexual orientation, nationality, disability status.
These attributes are represented with categorical variables in the data.
For example, we may have values {\it \{male, female, other \}} for gender, {\it \{European descent, African descent, $\dots$\}} for race, and so on.
Intersectional classes stem from the combination of protected attribute values.
For example, considering the attributes of gender and race alone, one intersectional class corresponds to individuals who are both female {\it and} of African descent, another to ones who are female {\it and} of European descent, and so on.
The number of intersectional classes grows exponentially with the number of protected attributes, which is one of the reasons why making a selection process fair with respect to admission rates of intersectional classes (or other criterion) can be computationally costly.

\spara{Candidate utility.}
We assume that every candidate is associated with a selection score that predicts future performance.
By default, the selection process aims to select those candidates who will have the better future performance at an institution -- and, from the institution's point of view, a candidate's selection score represents the {\it utility} of the candidate for the institution.
In what follows, we will write $u_{ij}$ to denote the utility of the $j^{\operatorname{th}}$ candidate from the $i^{\operatorname{th}}$ intersectional class.
Without loss of generality, we assume $u_{ij} \ge u_{ik}$ if $j < k$, i.e., candidates are sorted by non-increasing utility.


\spara{Selection rate.} The output of the selection process is a set of $k$ candidates. Assuming that there are $n$ candidates in total, this corresponds to a selection rate $\selectionrate = k/n$.
In what follows, we formulate our objective and operations in terms of $\selectionrate$ for ease of presentation, but we note that there is an exact correspondence with the number $k$ of selected candidates.
In other words, any policy based on selection rates can be transformed into a policy based on quotas, and viceversa.

\subsection{Problem formulation}
\label{sec:formulation}

In designing an affirmative action policy, we have two goals: first, to maximize the utility for the institution, i.e. to admit students with the best chances of having good academic performance; and second, to minimize disparities between the observed selection rates of different groups of candidates.
There is a natural trade-off between these two goals: to equalize selection rate across classes, we will have to forgo the selection of some candidates with high utility from an overrepresented class to enable the selection of candidates with lower utility from an underrepresented class.
Using the notation summarized on Table~\ref{tbl:notation} we formalize these goals and the trade-off between them into the objective function \objective below.

\begin{align}
\objective = \objective(\{\selected_{ij}\}) & = \totalutility(\{\selected_{ij}\})  - \weight \totaldiscrepancy(\{\selected_{ij}\})\label{eq:objective} \\
\totalutility(\setselected) & = \sum_{i=1}^{\nclasses} \sum_{j=1}^{\napplicants_j} \utility_{ij} \selected_{ij}\nonumber \\
\totaldiscrepancy(\setselected) & = \sum_{i=1}^{\nclasses} \left| \frac{\sum_{j=1}^{\napplicants_j} \selected_{ij}}{\napplicants_j} - \selectionrate \right|\nonumber
\end{align}
The objective \objective is expressed as a weighted sum of two terms, \totalutility and \totaldiscrepancy.
Essentially, \totalutility expresses the total utility from all selected candidates, while \totaldiscrepancy expresses the total absolute discrepancy of selection rates against the target selection rate \selectionrate across all intersectional classes.
%
%
The two quantities are combined into the same objective function via scalar factor \weight, which expresses the trade-off we are willing to make between utility, on one hand, and selection rate disparities across classes, on the other.
Essentially, $\weight$ expresses how many units of utility we are willing to forgo, to decrease disparity by one unit.
It is worth noting that this function penalizes both positive and negative disparities.
For instance, let assume that all members of one intersectional class are the only ones admitted. In this scenario, \totaldiscrepancy would be equal to $(1-\selectionrate) + (\nclasses-1)\selectionrate$.

We can now formalize the problem we wish to solve.

\begin{problem}
Given a pool of candidates divided into \nclasses non-overlapping intersectional classes, utilities $\{\utility_{ij}\}$, a target selection rate $\selectionrate$, and a trade-off factor \weight, make selections $\{\selected_{ij}\}$ of $\budget = \selectionrate\napplicants$ candidates so that $\objective(\{\selected_{ij}\})$ is maximized.
\label{problem:topk}
\end{problem}

%% file: 04-algorithms.tex

\section{Algorithms}
\label{sec:algorithms}

In this section we provide two algorithms to identify the set of applicants that maximize the objective function $\objective$ describred in Section~\ref{sec:formulation}: an optimal dynamic programming algorithm with running time $O(\nclasses \cdot \napplicants^2 )$, and a faster greedy heuristic that requires $O(\nclasses \cdot \napplicants)$ operations.
Naturally, if $\weight = 0$, the optimal solution corresponds to the top $\budget = \selectionrate \cdot \napplicants$ of students in the admission ranking, regardless of their protected attributes.
If, on the contrary, we set $\weight$ to a sufficiently large value, we obtain a solution with minimum discrepancies, composed of the top $\selectionrate \cdot \napplicants_j$ applicants from each intersectional class $j$.
We remark that in all solutions generated by these algorithms the admitted applicants are the top-scoring \emph{within} each class.
The goal of all algorithms is to determine the decisions \setselected; all other quantities of Table~\ref{tbl:notation} are considered given and fixed.


\subsection{Optimal Dynamic Programming Approach}

The \gls{dp} maintains a table $T$ having
$\nclasses$ rows (one for each intersectional class), and
$\budget = \napplicants \cdot \selectionrate$ columns (one for each vacancy).
This table holds in cell $T(i,j)$ the maximum value attainable for the objective function when admitting a total of $j$ applicants exclusively from intersectional classes $1, 2, \dots, i$. The final value of the objective function is observed in cell $T(\nclasses, \budget)$.
%

\gls{dp} maintains a second, auxiliary table $\auxtable$ of $i$ rows and $j$ columns, which contains in cell $\auxtable(i,j)$ the objective function contribution for intersectional class $i$ if the top $j$ students from this intersectional class are admitted in the final solution. Formally:
\begin{equation}
\label{eq:aux_table}
R(i,j) = \sum_{k=1}^j \utility_{ik} - \weight \left| \frac{j}{\napplicants_i} - \selectionrate \right|~.
\end{equation}
\gls{dp} is descibed in Algorithm~\ref{alg:dynamic}, and it outputs the maximum value of the objective number.
The algorithm performs careful book-keeping so as to never admit from one class more than its $\napplicants_j$ candidates.
Specifically, updates are performed only for up to $\min\{\napplicants_j, \budget = \napplicants \cdot \selectionrate\}$ candidates from class $j$ (Alg.~\ref{alg:dynamic}, for-loop at line~\ref{alg1-loop2}) but not for more ((Alg.~\ref{alg:dynamic}, for-loop at line-\ref{alg1-loop3-end}).
To retrieve the number of admitted students per class, which unambiguously identifies the admitted students, we backtrack using an additional data structure that maintains the maximum index found on line~\ref{alg1-update} of the algorithm; details are omitted for presentation clarity.
An implementation of this algorithm can be found in our code and data release.\footnote{URL omitted for double-blind review.}

\begin{algorithm}[H]
    \caption{\gls{dp}}
    \label{alg:dynamic}

\begin{algorithmic}[1]
	\STATE Initialize $U$ using Equation~\ref{eq:aux_table}
	\STATE Initialize $T \leftarrow 0$ 
  \STATE $T(-1, \cdot) \leftarrow T(\cdot, -1) \leftarrow 0$
	\FOR{\label{alg1-loop1}$i = 1, \dots, \nclasses$}
		\FOR{\label{alg1-loop2}$j = 1, \dots, \min\{\napplicants_j, \napplicants \cdot \selectionrate\}$}
			\STATE\label{alg1-update} $T_{ij} = \max_{k = 0, 1, \dots, j} \left[ U(i, k) + T(i-1, j-k) \right]$
		\ENDFOR
    \FOR{\label{alg1-loop3}$j = \min\{\napplicants_j, \napplicants \cdot \selectionrate\} + 1, \dots, \napplicants \cdot \selectionrate$}
      \STATE $T_{ij} = T_{i(j-1)}$
    \ENDFOR\label{alg1-loop3-end}
	\ENDFOR
	\RETURN $T(\nclasses, \napplicants \cdot \selectionrate)$
\end{algorithmic}
\end{algorithm}

\spara{Analysis.} By observing the nested loops in the algorithm in lines \ref{alg1-loop1}-\ref{alg1-update}, we can see its running time is $O(\nclasses \cdot \napplicants^2)$.
Its optimality can be proven by induction on the number of classes $\nclasses$. If $\nclasses=1$ then the algorithm simply picks the number of students $j$ that maximizes $U(1,j)$, which corresponds to the optimal value of $\objective$, which is what we seek to optimize (compare equations \ref{eq:objective} and \ref{eq:aux_table}).
Now let us assume \gls{dp} yields the optimal solution with $i$ classes. If we consider one additional class, we will consider for each number of candidates $j = 1, 2, \dots, M = \min\{\napplicants_j, \napplicants \cdot \selectionrate\}$ in the additional class (line \ref{alg1-loop2}), the best solution of $M - j$ candidates from the the first $i$ classes.
This is optimal as per the inductive hypothesis, due to the additivity of the objective function \objective (Eq.~\ref{eq:objective}) over disjoint classes. Hence, the output is optimal.

\subsection{Greedy Approximation}

The \gls{greedy}, described  in pseudocode in Algorithm~\ref{alg:greedy}, proceeds as follows: at every step it adds the applicant who increases the value of the objective $\objective$ the most.
%
%
The algorithm runs until $\budget = \napplicants \cdot \selectionrate$ applicants are admitted.

\begin{algorithm}[H]
    \caption{\gls{greedy}}
    \label{alg:greedy}
\begin{algorithmic}[1]
 	\STATE Selected $\leftarrow \emptyset$
	\FOR{$j = 1, \dots, \napplicants \cdot \selectionrate$}
    \STATE /* Select the candidate that maximizes \objective  */
    \STATE $x \leftarrow \operatorname{argmax}_{x \not\in \operatorname{Selected}} \objective(\operatorname{Selected} \cup x)$
    \STATE Selected $\leftarrow$ Selected $\cup~~x$
	\ENDFOR
	\RETURN Selected
\end{algorithmic}
\end{algorithm}

\spara{Analysis.} The running time of the algorithm is $O(\budget\napplicants)$.
Even though empirically we find that the greedy algorithm often produces the same solution as the optimal algorithm for our data, there are cases in which it produces a sub-optimal solution (by a small margin).
We elaborate on this point in Section~\ref{sec:results}.

Note also that the running time of \gls{greedy} could be further optimized.
To see why, notice that \gls{greedy} always selects candidates in decreasing utility order within class.
Therefore, the acceptance of each candidate has an effect to the objective function \objective that is known in advance, allowing us to avoid the expensive step (4) in Alg.~\ref{alg:greedy}.
In more detail, we calculate the benefit each of the top-\budget candidates brings to \objective when added by greedy; the running time of this operation is $O(\classes\budget)$.
Subsequently, the top-\budget candidates of each class are traversed in parallel and ``merged'' to keep the top-\budget elements from all classes; the running time of this operation is $O(\budget)$.
Therefore, greedy can be optimized to run in $O(\classes\budget) < O(\napplicants\budget)$ time.

%% file: 05-experiments.tex

\section{Experimental Setting}
\label{sec:experiments}

We perform several experiments in the context of university admissions, using a large dataset of university applicants in an OECD country, and measuring the effectiveness and efficiency of the algorithms we have proposed.
In terms of the setting described in Section~\ref{sec:statement}, each university applicant is considered a {\it candidate} in our problem formulation; {\it intersectional classes} are defined based on attributes associated with the applicants; the program that accepts applications is the {\it institution} that selects some of them for admission; and the admission of the best applicants for a university program is framed as a problem of {\it top-\budget candidate selection} (Problem~\ref{problem:topk}).

\subsection{Dataset Description}
\label{sec:datasets}

Undergraduate programs in the studied country admit students based on a linear combination of standardized test scores across four different subjects (math, language, natural sciences, and social sciences) and their high school grades.
In 2017, the most recent year for which we have an entire anonymized dataset, over 300,000 people registered to take the standardized tests needed for the application process.
Out of these, only about 120,000 actually took the tests and applied to one or more programs.
A student's application consists of a ranked list of up to 10 programs in decreasing order of preference.
Of the 120,000 applicants, 75,000 were eventually admitted to some program; except for students with exceptionally high scores and grades, the program to which they were admitted was not their top preference.
For all the candidates, we were given access to demographic, socio-economic and educational information including: nationality, gender, birth year, civil status, high school of provenance (type and region), and household characteristics (income decile, parents' occupation and level of education). High school grades and standardized test results were provided as well.
Other researchers can apply to access the same dataset.\footnote{Instructions will be provided in our camera-ready version.}

Each program weighs scores and high school grades differently, and can set minimum thresholds of scores for applicants.
In general, admission scores of successful applications are above 450 points, with more prestigious and competitive programs usually accepting students above 750 points, out of a maximum of 850.

It is worth noting that not just the chance of getting admitted to university changes for different social groups, but access to quality education itself is strongly correlated with socio-economic status.
Certain sections of society that make up a large portion of the total population are scarcely represented in the application process.
For instance, many students coming from families that lack the means to support them through university, may not even apply to a tertiary education process.
%

\subsection{Experiments}
\label{subsec:experiments}

Our experiments consider the 75,000 
students who were eventually admitted to a program.
Since the majority of programs received fewer applicants than the number of vacancies they offered,
and students can apply to up to 10 programs,
this means we removed from the dataset students who either have very low scores, such that they would not qualify for most of the programs; or for some reason 
did not apply to enough programs to be admitted by at least one.
For the intersectional classes we consider the intersection of the $3 \times 2 \times 2$ categories described in Table~\ref{tbl:coding}, generating 12 groups that differ significantly in their admission score distributions.

\input{figs-utility-vs-discrepancy-single}

We conducted two different sets of experiments to which we refer as \singletrack and \separatetrack, described below.

\spara{\singletrack.}
In the first set of experiments, we consider the hypothetical scenario in which all students in our data applied to the same one program.
The purpose of this is to explore the effect of the algorithms in a setting that is independent of their particular choice of program.
For both the greedy and the dynamic approaches, we ran algorithms with four different admission rates ($p \in \{ 0.05, 0.15, 0.30, 0.50 \}$), gradually increasing the weight $\weight$ from zero until we reach no discrepancy.

\spara{\separatetrack.}
In the second set of experiments, we tested our algorithms on a few specific programs.
We chose four programs among the most selective ones (these are four medical schools), and other four among those that had the highest number of admitted students (these are two engineering schools and two business schools).
This second set of experiments explores the effect of the algorithms in more realistic settings -- and helps highlight how the same approach performs across different programs.

%% file: figs-utility-vs-discrepancy-single.tex

\begin{figure*}[t]

  \hspace{-1.8em}
    \includegraphics[trim=0 0 0 0, clip, width=.30\textwidth]{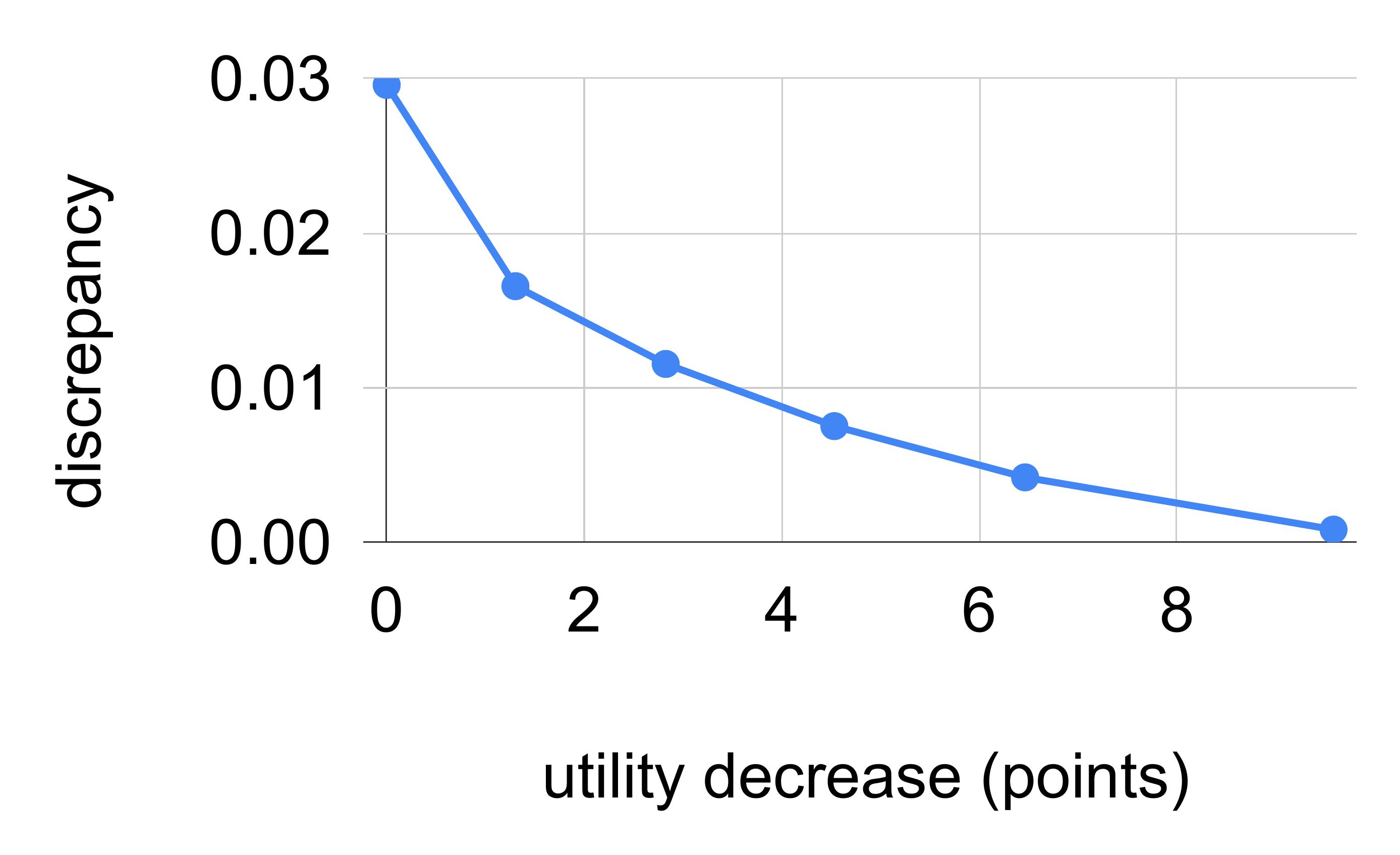}~\includegraphics[trim=0 0 0 0, clip, width=.30\textwidth]{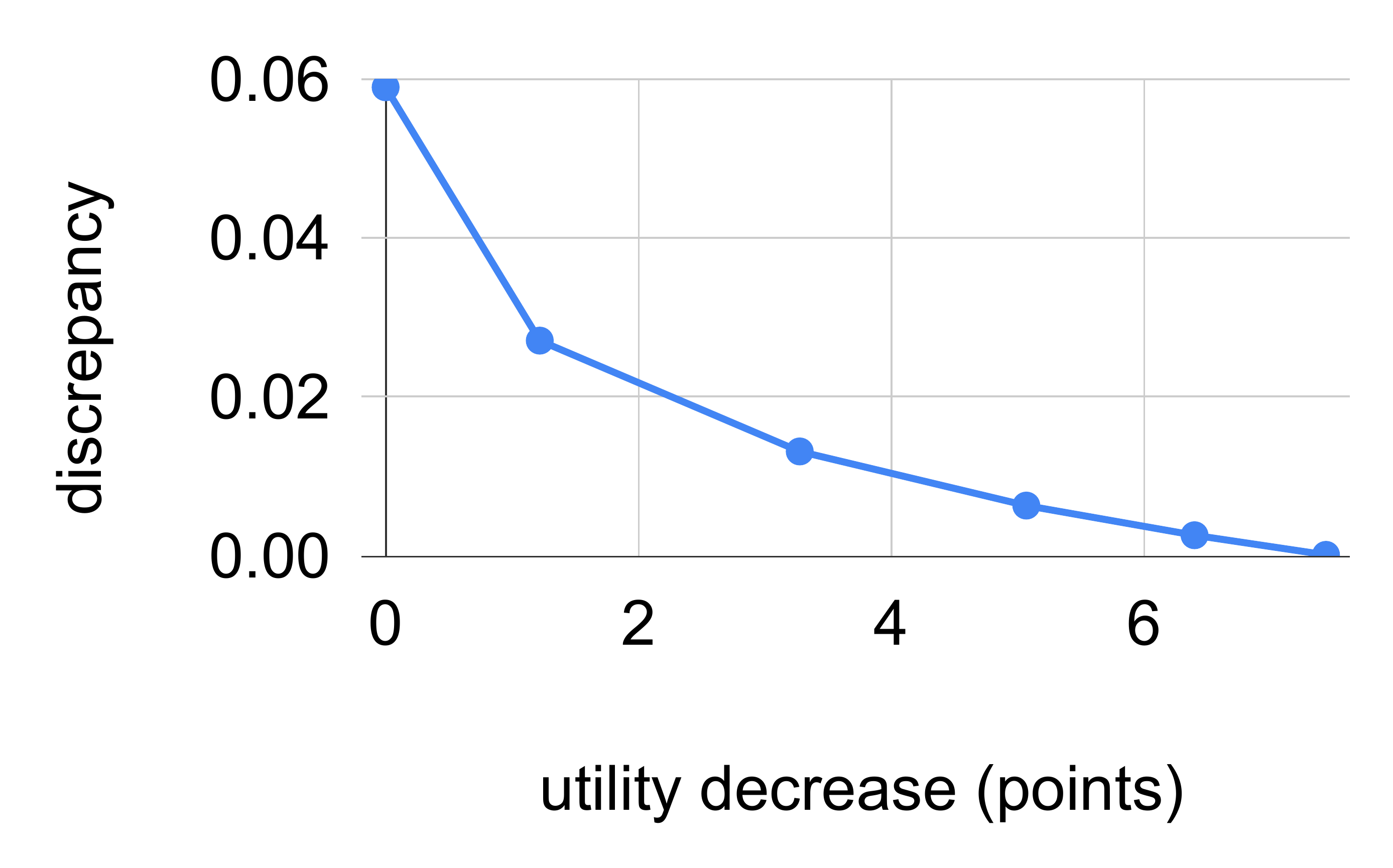}~\includegraphics[trim=0 0 0 0, clip, width=.30\textwidth]{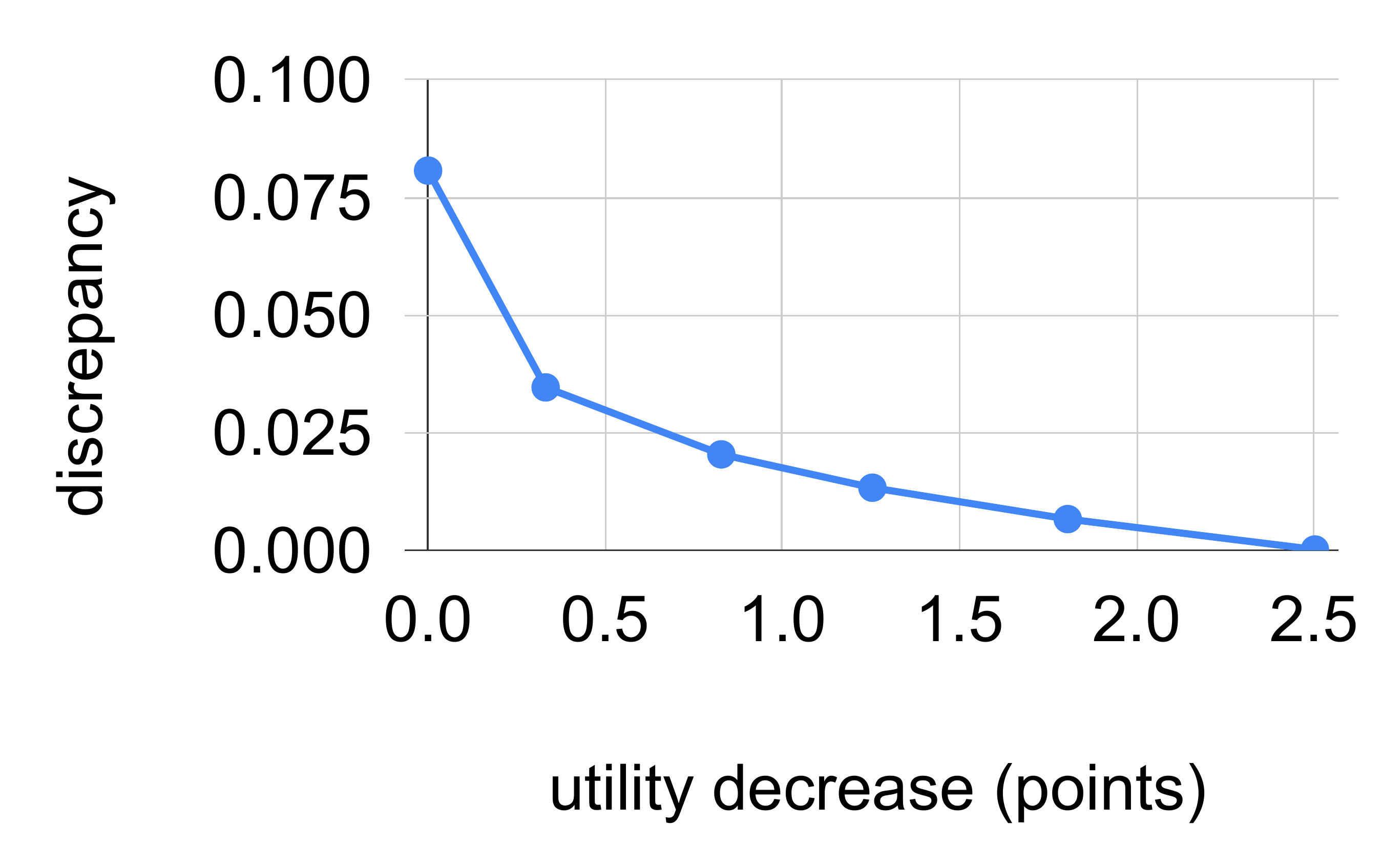}

  \subfloat[$\selectionrate=0.05$]{
  \hspace{-1.8em}
    \includegraphics[trim=0 0 0 0, clip, width=.30\textwidth]{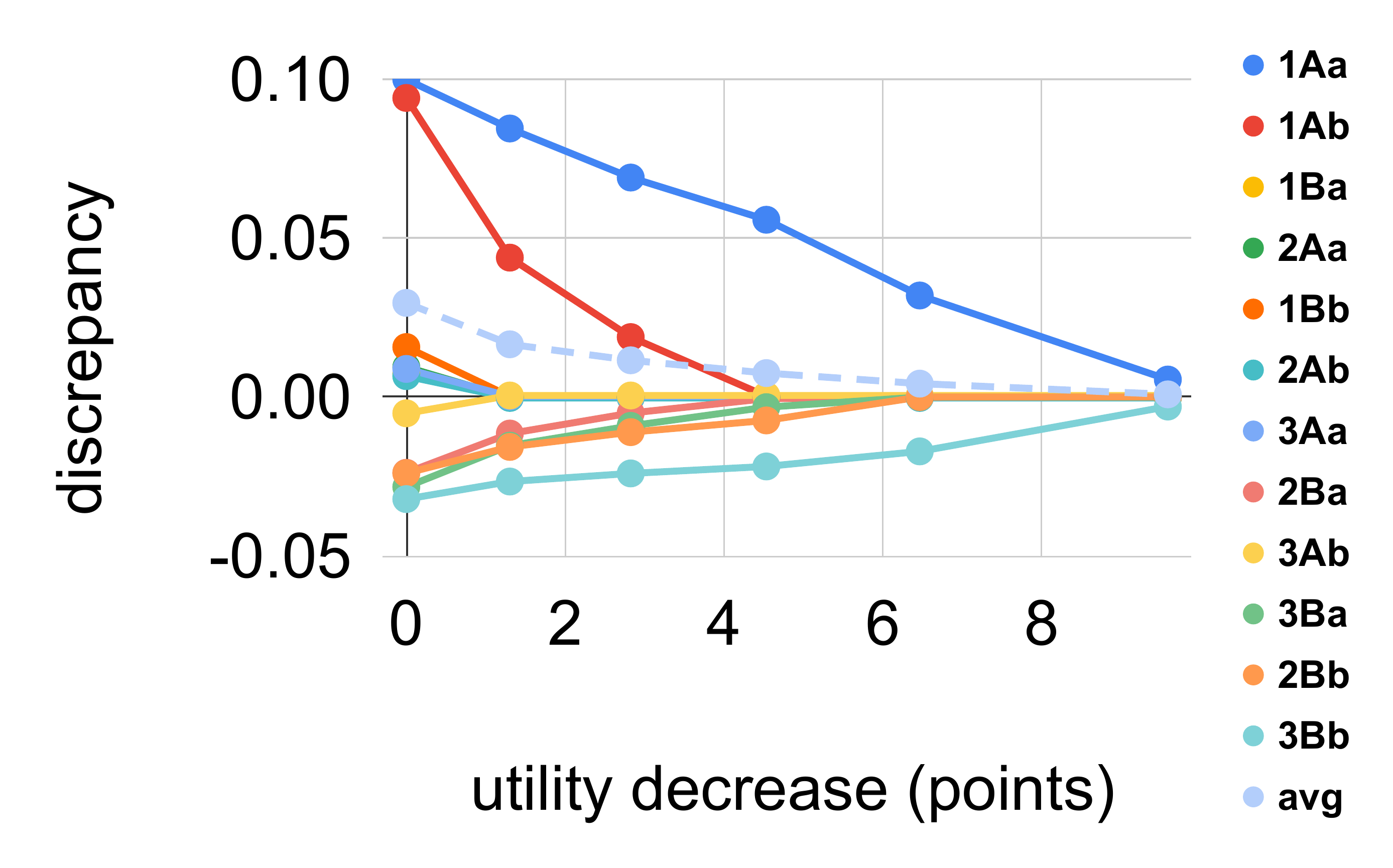}
  }
  \subfloat[$\selectionrate=0.15$]{
    \includegraphics[trim=0 0 0 0, clip, width=.30\textwidth]{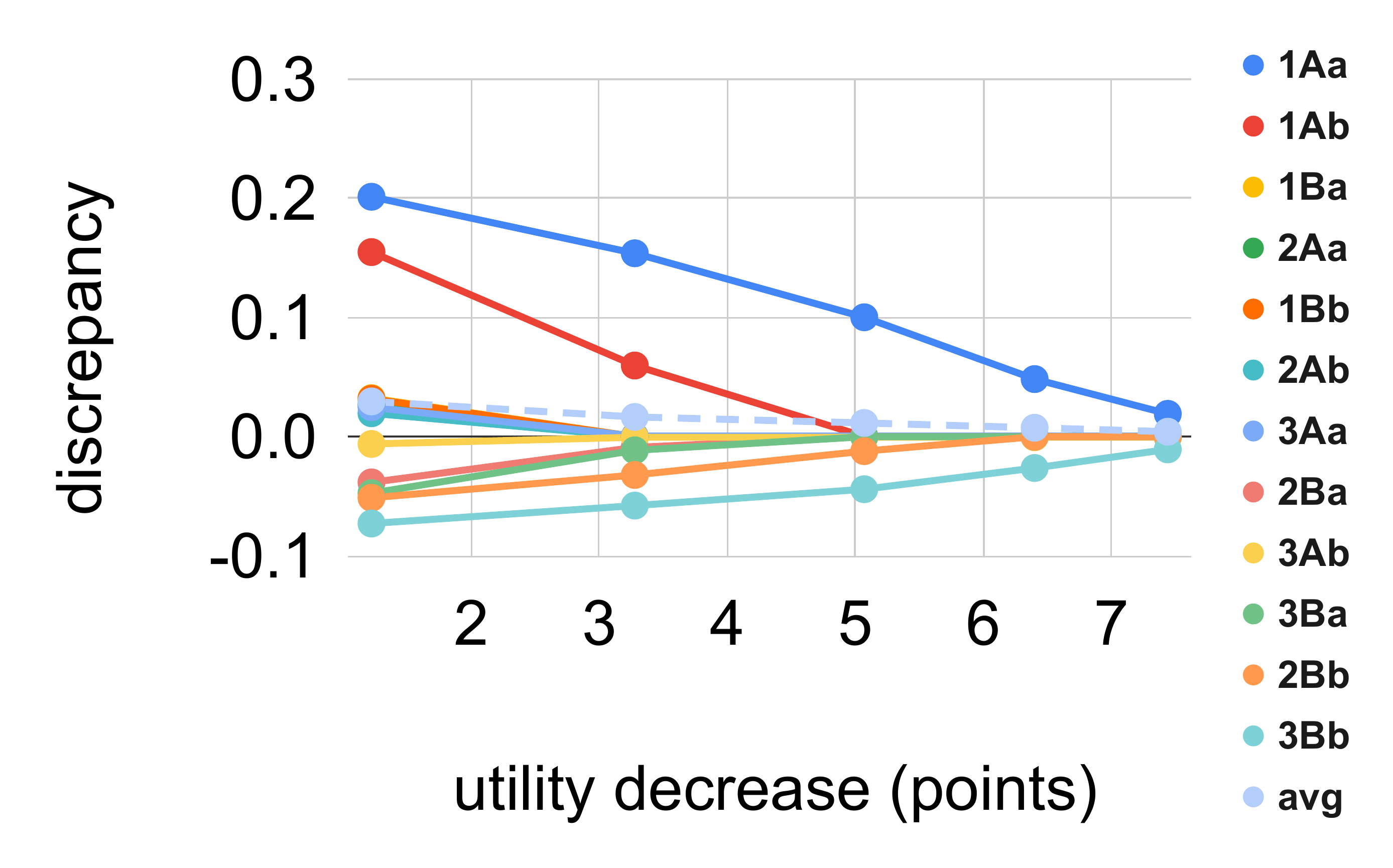}
  }
  \subfloat[$\selectionrate=0.50$]{
    \includegraphics[trim=0 0 0 0, clip, width=.30\textwidth]{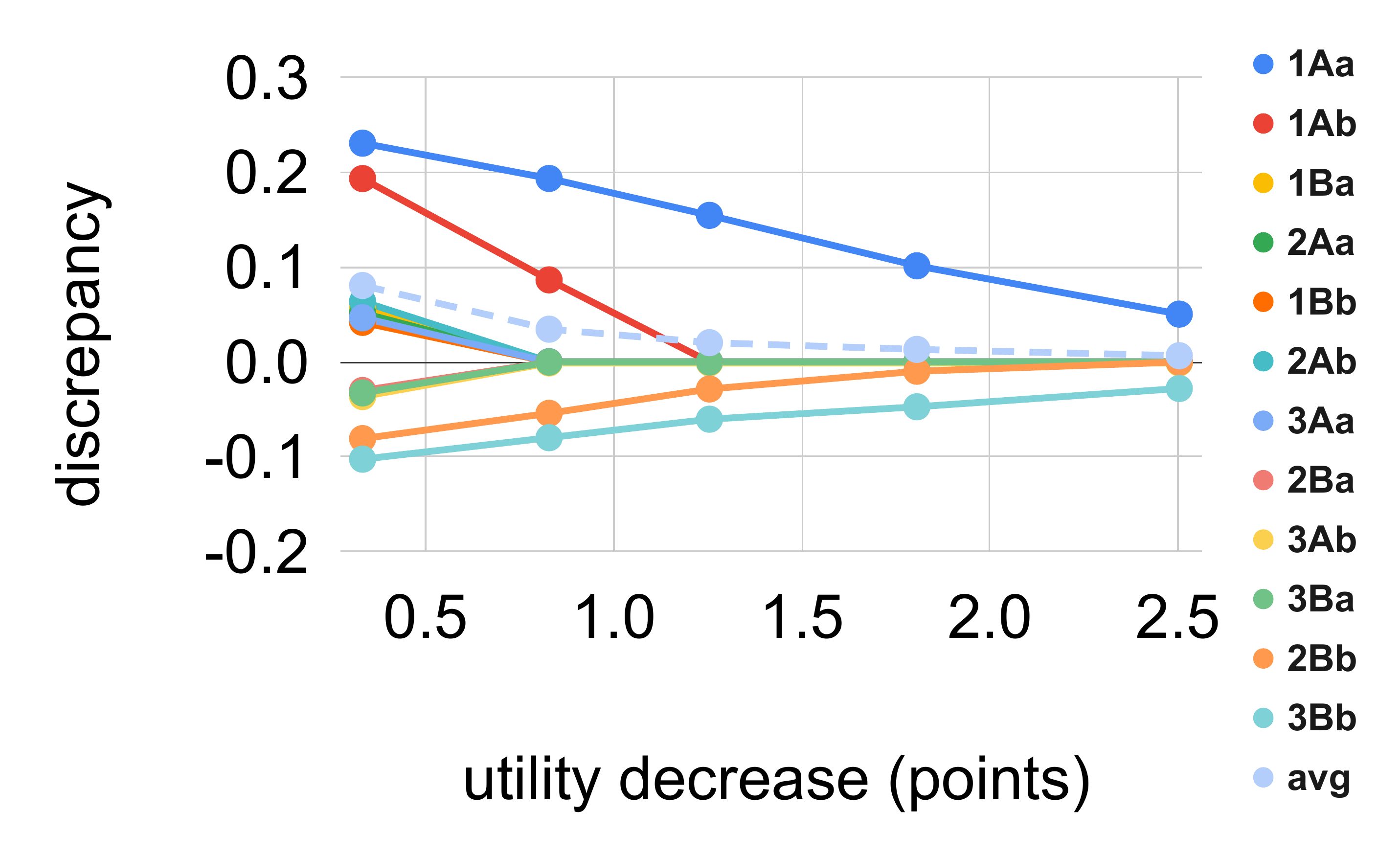}
  }

  \caption{Average (top) and per-group (bottom) discrepancies in admission rates if all students were to apply to the same program, for four different target admission rates.
  In this and the following figures, larger values of $\weight$ lead to smaller discrepancies and less utility.}
  \label{fig:discrepancies}
\end{figure*}

%% file: 06-results.tex

\input{figs-utility-vs-discrepancy-separate}

\section{Results}
\label{sec:results}

We now present the results from the experimental evaluation.
A first observation is that in practice achieving parity of acceptance rate between the intersectional classes leads to a decrease of just a few points in the average utility, i.e., the average admission score of accepted applicants.
A second observation is that even if the greedy algorithm is not guaranteed to yield an optimal solution, it almost always finds the same solution as the dynamic programming approach and it runs much faster.
In what follows, the reported results are obtained with the \gls{dp} algorithm, unless explicitly stated otherwise.

\subsection{Effectiveness}

\spara{\singletrack.}
Figure~\ref{fig:discrepancies} shows results for the hypothetical setting where all students in the dataset apply for the same program.
In the top of Figure~\ref{fig:discrepancies}, we plot the average discrepancy in admission rates across all the 12 intersectional groups, for three different target admission rates, and for varying values of the parameter $\weight$.
The left-most points in these plots (i.e., for $utility\ decrease = 0$) correspond to the case of $\weight = 0$, i.e., when no affirmative action is enacted.
Moreover, the right-most points in the same plots correspond to the case when the discrepancy of admission rates is practically removed (specifically, when \totaldiscrepancy falls below $0.01$).
We can clearly see a trade-off between disparity and average utility (i.e., average admission score of admitted applicants).
Specifically, there is a drop of about $10$ utility points to reach perfect admission rate parity - when the target admission rate is the lowest among those tested ($p=0.05$).

Notice that, as the admission rate increases (from $p=0.05$ on the left-most plot to $p=0.50$ to the right-most one), the utility decrease required to reach parity decreases as well (from about $10$ utility points, to $2.5$ utility points).
This is due to the fact that inequalities are more pronounced at the higher end of utility, i.e., for applicants with the best scores.
That is, the score distributions of different intersectional classes are more different in the top tail.
We remark that in a practical application, acceptance rate parity might not be the goal, given that even a partial reduction of a large historical disparity could be sufficient and desirable as policy objective.

In the bottom of Figure~\ref{fig:discrepancies}, we break down the discrepancy across intersectional classes.
As expected, we can see that an increase in $\weight$ leads to admitting more students from the disadvantaged classes and fewer students from the privileged classes.

How would someone use these plots to design an affirmative action policy?
One natural way is the following:
given a target selection rate \selectionrate, the policy maker would use the plot corresponding to it; and given an acceptable decrease in utility $\delta_\totalutility$, the policy maker would choose the value for $\weight$ that corresponds to the point in the plot with $utility\ decrease = \delta_\totalutility$.

\begin{figure}[t]
\includegraphics[trim=0 0 0 0, clip, width=0.8\columnwidth]{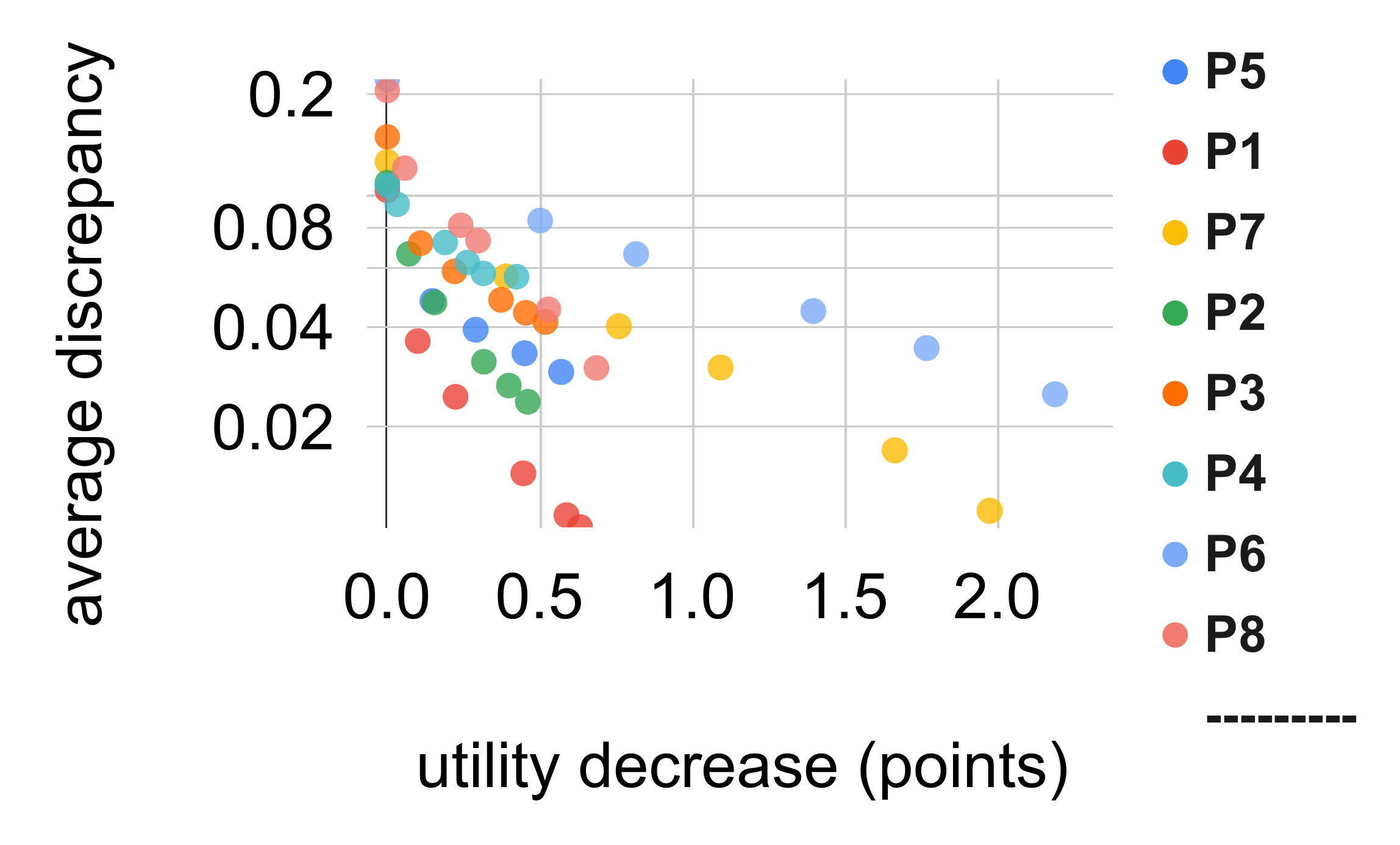}
\caption{(Best seen in color.) Per-program decrease in utility as average discrepancies of admission rates are reduced, for eight different programs.
Each color corresponds to a different program, and different points correspond to different values of the weight $\lambda$, with larger weights leading to less discrepancy and less utility.
Note the y scale is logarithmic.}
  \label{fig:singleprogramsummary}
\end{figure}

\spara{\separatetrack.}
Figure~\ref{fig:per-program-performance} shows results for affirmative action policies applied on one of eight individual programs.
In this setting, the applicant pool consists of those applicants in our data who, in their application,  ranked each program higher than the one where they eventually got admitted.

A first caveat in this applied analysis case is the often small number of applicants in some of the 12 intersectional classes.
In fact, for highly competitive programs, some of the most disadvantaged groups are practically absent in the application pool.
In practice, this means that the discrepancy values for intersectional classes vary abruptly with the addition of admitted applicants, and the results cannot be considered robust.
For instance, when an intersectional class contains a single applicant, not admitting him/her would lead to $-p$ discrepancy for that class, or $1-p$ otherwise.
For this reason, we adapted our objective function (and therefore, also the algorithms) to ignore intersectional classes containing fewer than three ($3$) applicants.
Even with this adaptation, however, the remaining intersectional classes are sufficient to obtain meaningful insights from the application of our approach on a variety of real situations.

As shown in Figure~\ref{fig:singleprogramsummary}, the general trend is similar for all individual programs, with a steep decrease in average discrepancy (note the y scale is logarithmic) as the average utility decreases by one or two points.
Unlike the hypothetical \singletrack case (Figure~\ref{fig:discrepancies}), individual programs require a comparatively smaller sacrifice of utility to reach parity, also because generally they show less inter-class discrepancy (Figure~ \ref{fig:per-program-performance}).

Arguably, in a program admitting students with 750 out of 850 points of score in standardized tests, a difference of one or two points (which may represent a difference of less than one question answered incorrectly on a test) cannot be expected to have a noticeable effect on their academic performance or their chances of graduating several years later.
Moreover, as particularly evident in figures~\ref{fig:per-program-performance}(d) and ~\ref{fig:per-program-performance}(h), when algorithms are applied in a single program scenario, it is quite common for most of the intersectional classes to maintain the same discrepancy throughout the entire process.
Usually, only the most privileged and most disadvantaged groups, such as 1Aa, 1Ab, 1Ba, 1Bb, 2Ba, and 3Ba, see their discrepancy and average admission score modified in order to reach parity.

\subsection{Efficiency}

\begin{figure}[t]
\centering\includegraphics[trim=0 0 0 0, clip,width=1\columnwidth]{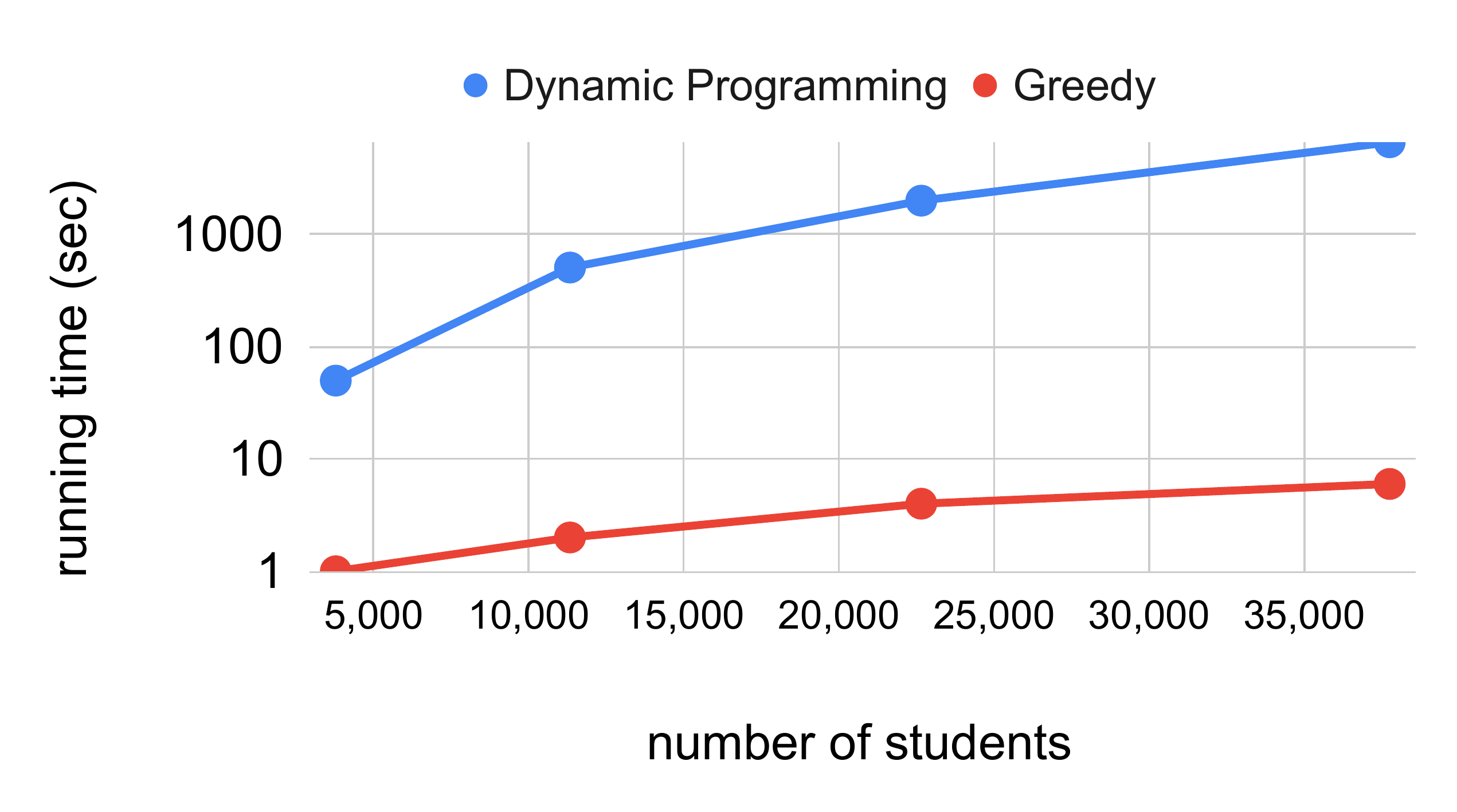}
\caption{Comparison of running time of the dynamic programming algorithm and the greedy algorithm.}
\label{fig:efficiency}
\end{figure}

Figure~\ref{fig:efficiency} plots the running time of the algorithm for sub-samples of increasing size drawn from the total population of students.
We observe that the greedy method is orders of magnitude faster than the dynamic programming method.
%
In practice, throughout all of our experiments, the greedy algorithm only failed to find the optimum result in one configuration: the \singletrack setting with admission rate $0.30$ and $\weight=0$.
However, in the case of university admission and other high-risk scenarios affecting the future of thousands of people, we recommend to use the dynamic programming algorithm that has an optimality guarantee.

%% file: figs-utility-vs-discrepancy-separate.tex

\begin{figure*}[t]
  \subfloat[P1, $n=156, \overline{u}=796, \selectionrate=0.20$]{
  \hspace{-1.1em}
    \includegraphics[trim=0 0 0 0, clip, width=.66\columnwidth]{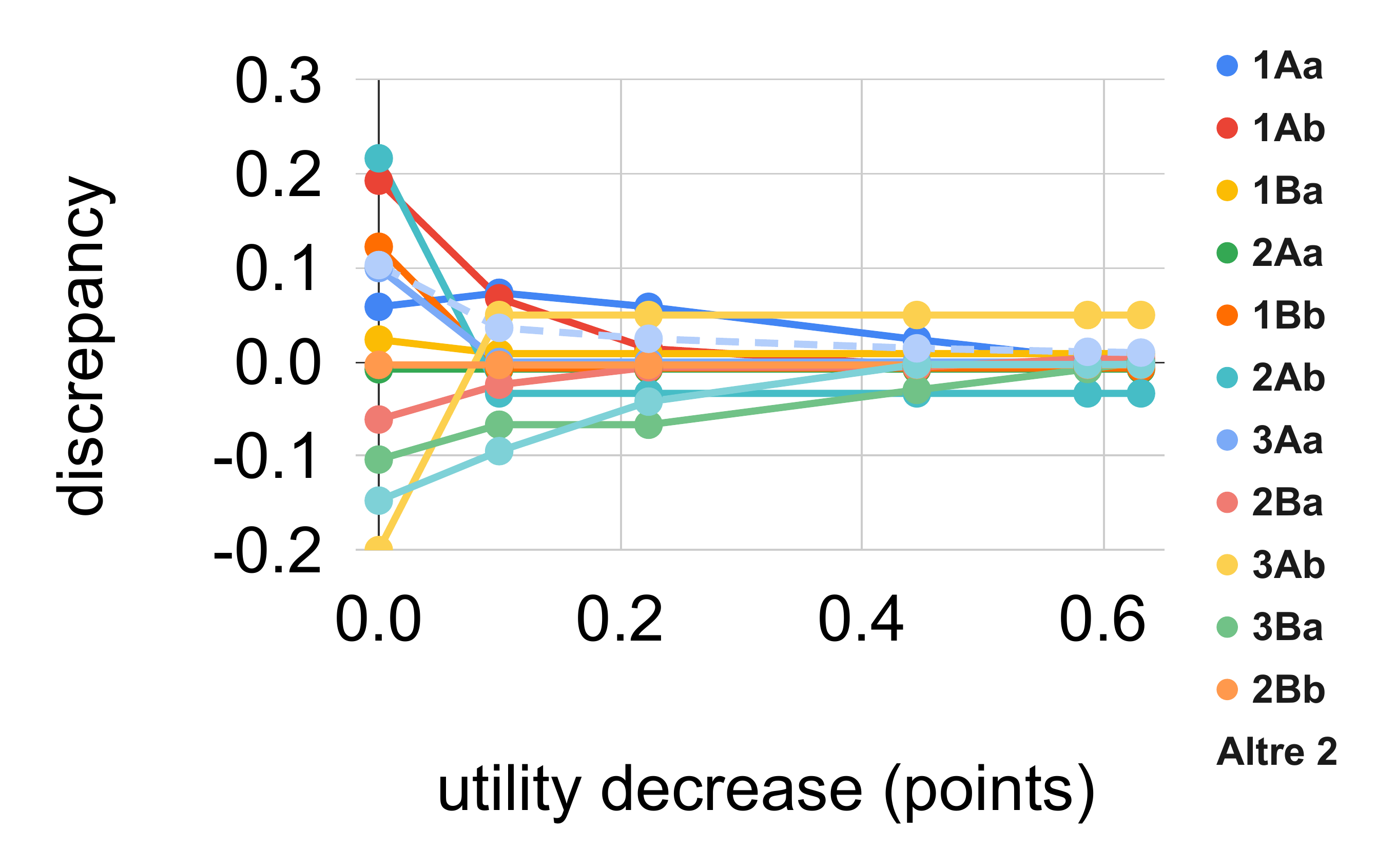}
  }
  \subfloat[P2, $n=82, \overline{u}=804, \selectionrate=0.18$]{
    \includegraphics[trim=0 0 0 0, clip, width=.66\columnwidth]{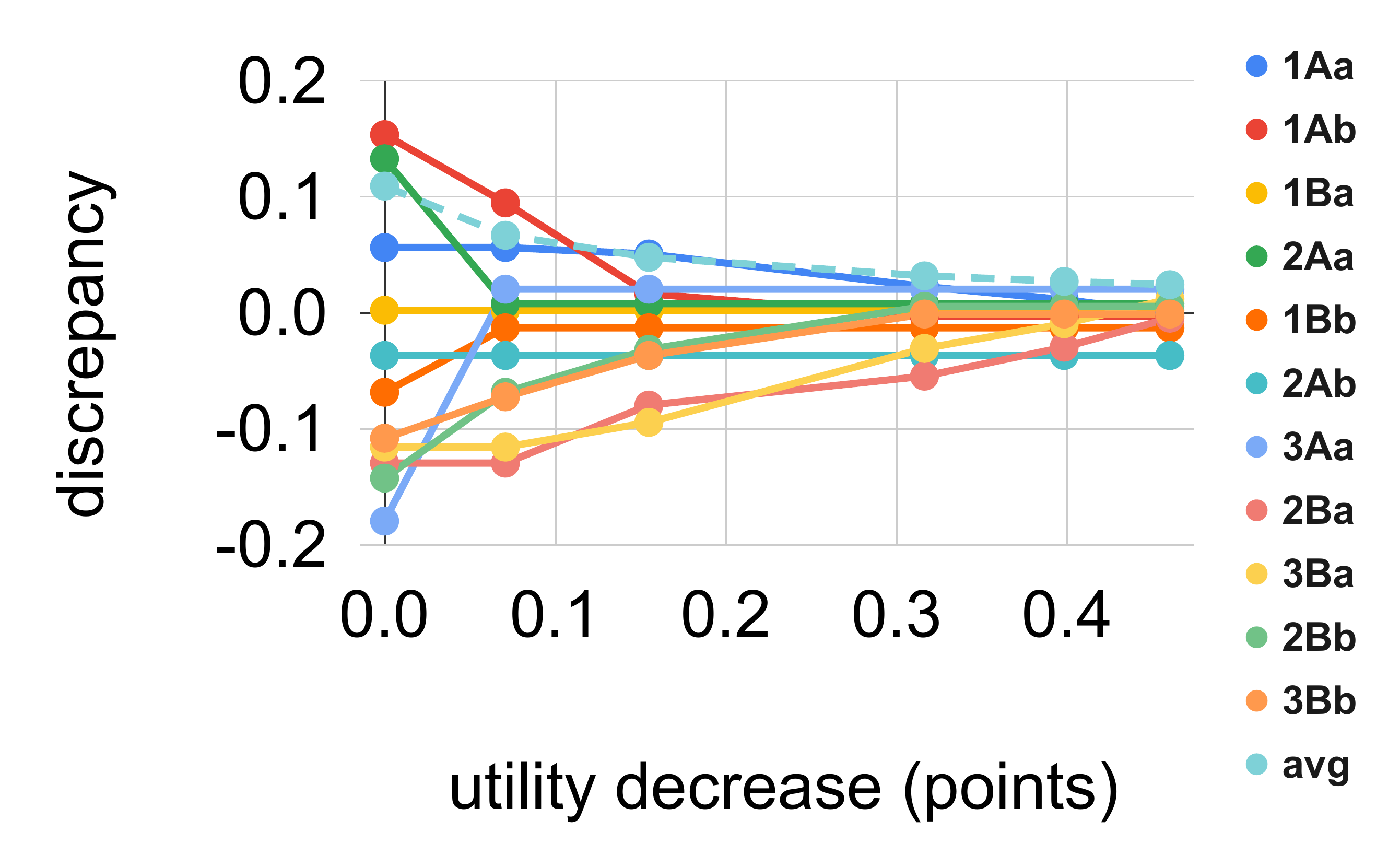}
  }
  \subfloat[P3, $n=105, \overline{u}=782, \selectionrate=0.18$]{
    \includegraphics[trim=0 0 0 0, clip, width=.66\columnwidth]{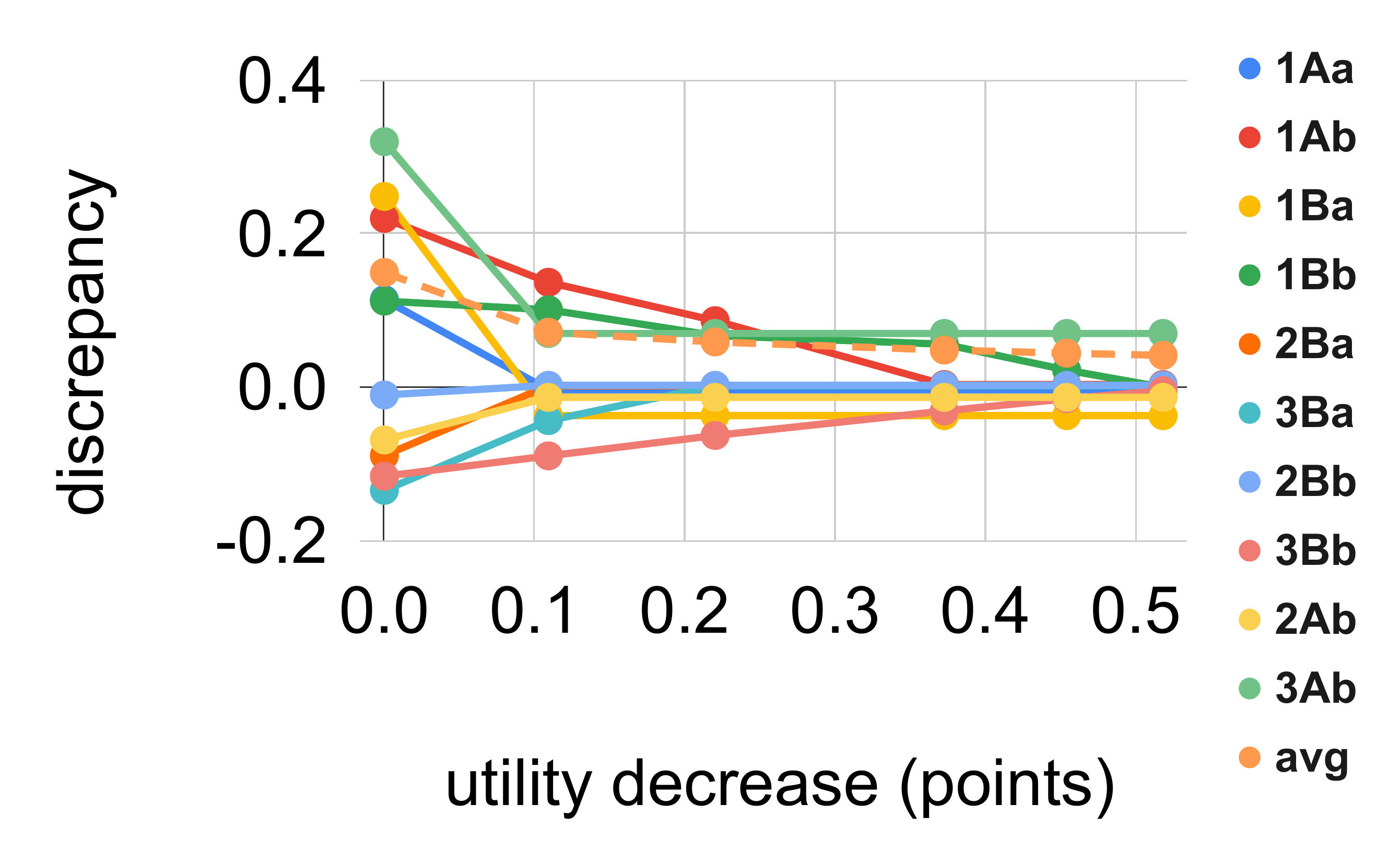}
  }

  \subfloat[P4, $n=42, \overline{u}=761, \selectionrate=0.17$]{
    \includegraphics[trim=0 0 0 0, clip, width=.66\columnwidth]{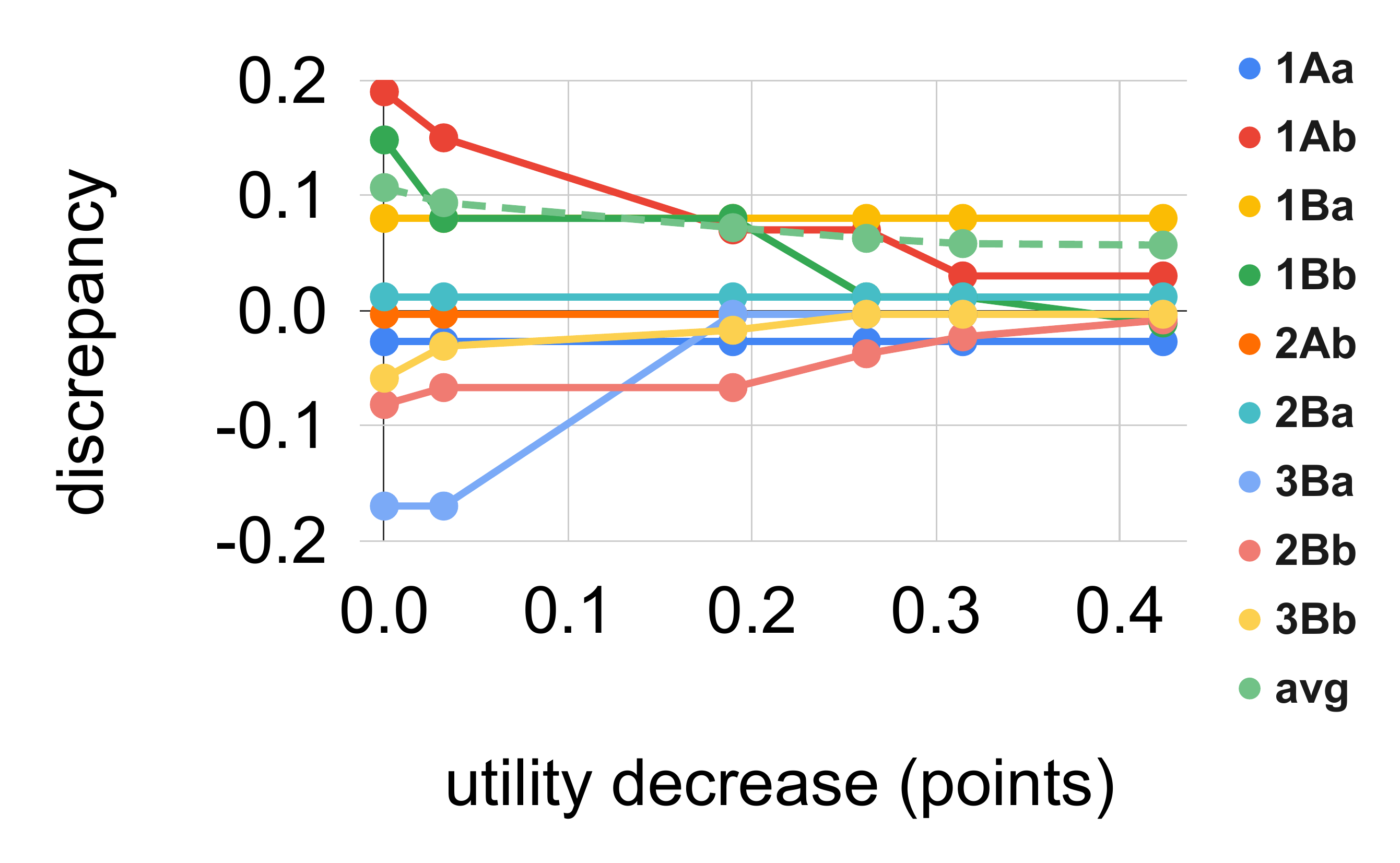}
  }
  \subfloat[P5, $n=687, \overline{u}=747, \selectionrate=0.39$]{
  \hspace{-1.1em}
    \includegraphics[trim=0 0 0 0, clip, width=.66\columnwidth]{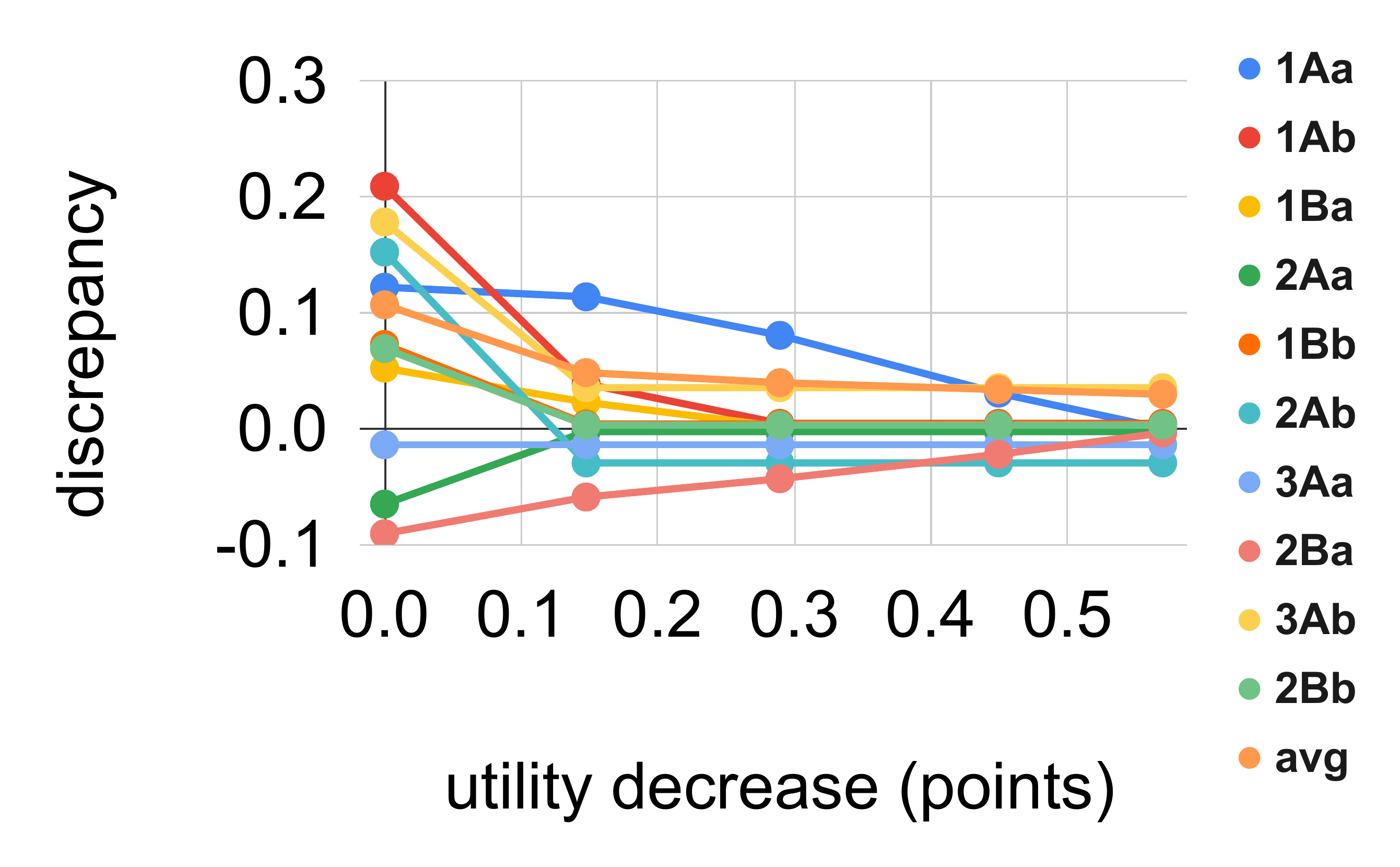}
  }
  \subfloat[P6, $n=367, \overline{u}=741, \selectionrate=0.53$]{
    \includegraphics[trim=0 0 0 0, clip, width=.66\columnwidth]{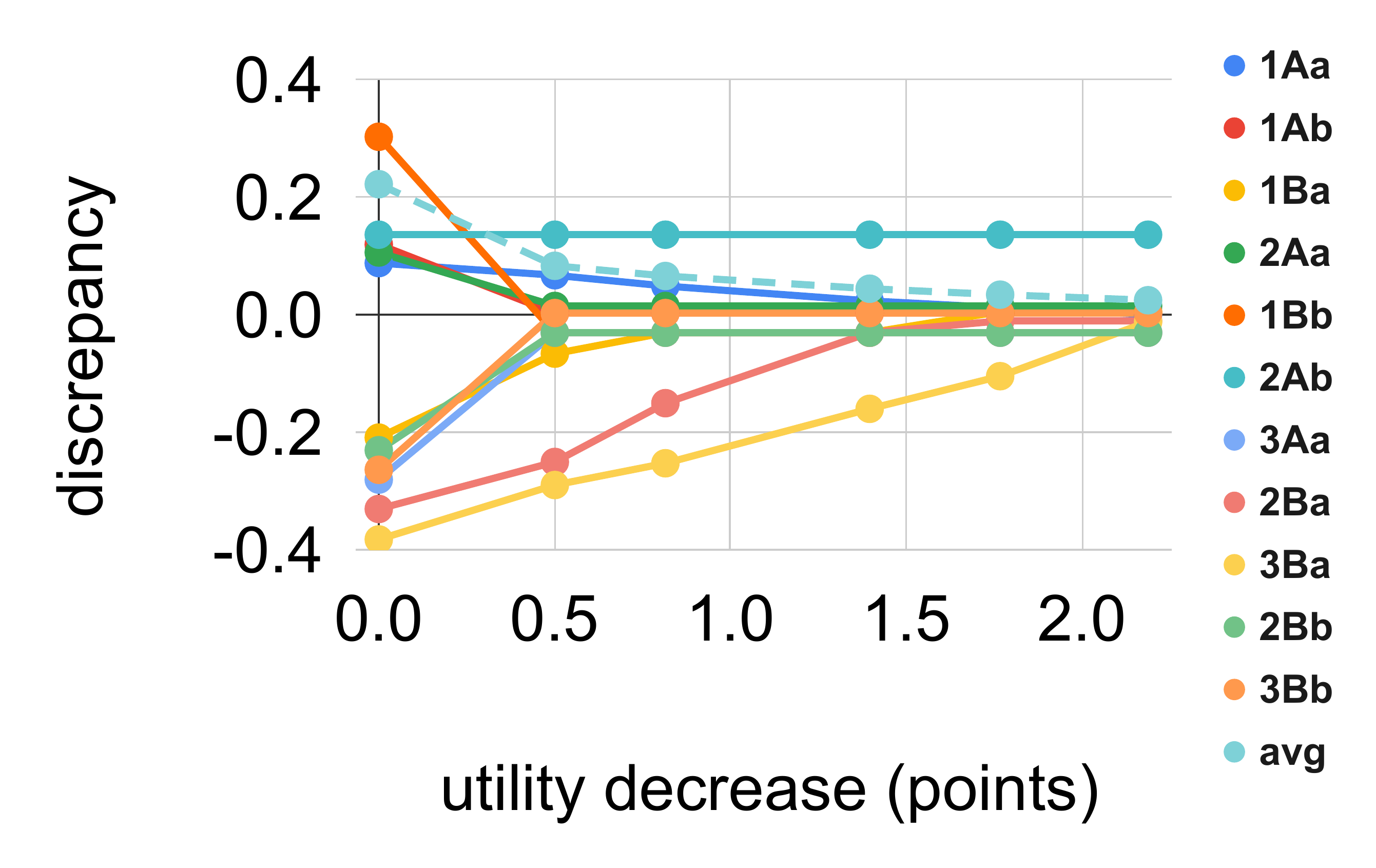}
  }

  \subfloat[P7, $n=603, \overline{u}=764, \selectionrate=0.51$]{
  \hspace{-1.1em}
    \includegraphics[trim=0 0 0 0, clip, width=.66\columnwidth]{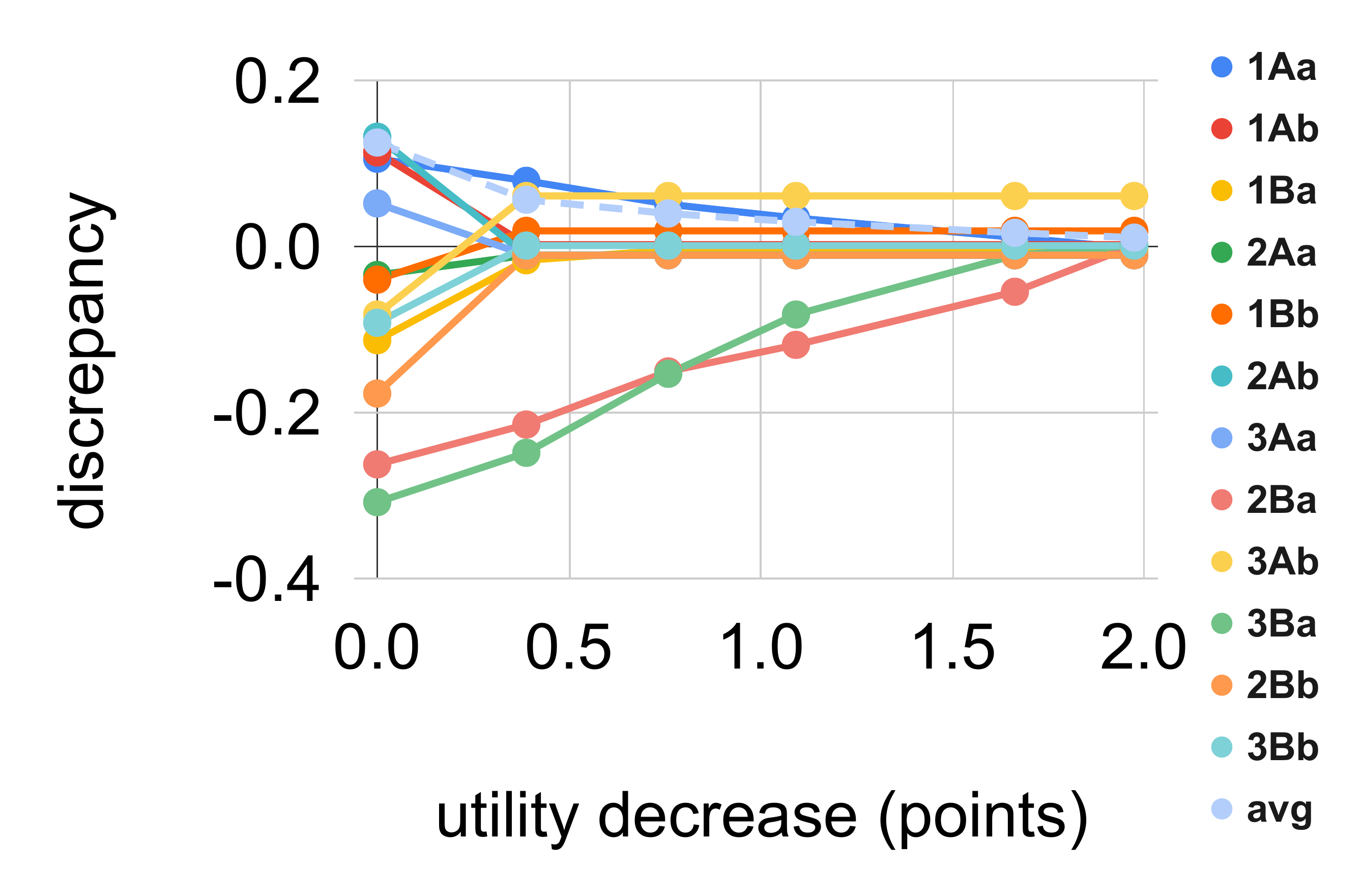}
  }
  \subfloat[P8, $n=490, \overline{u}=669, \selectionrate=0.70$]{
    \includegraphics[trim=0 0 0 0, clip, width=.66\columnwidth]{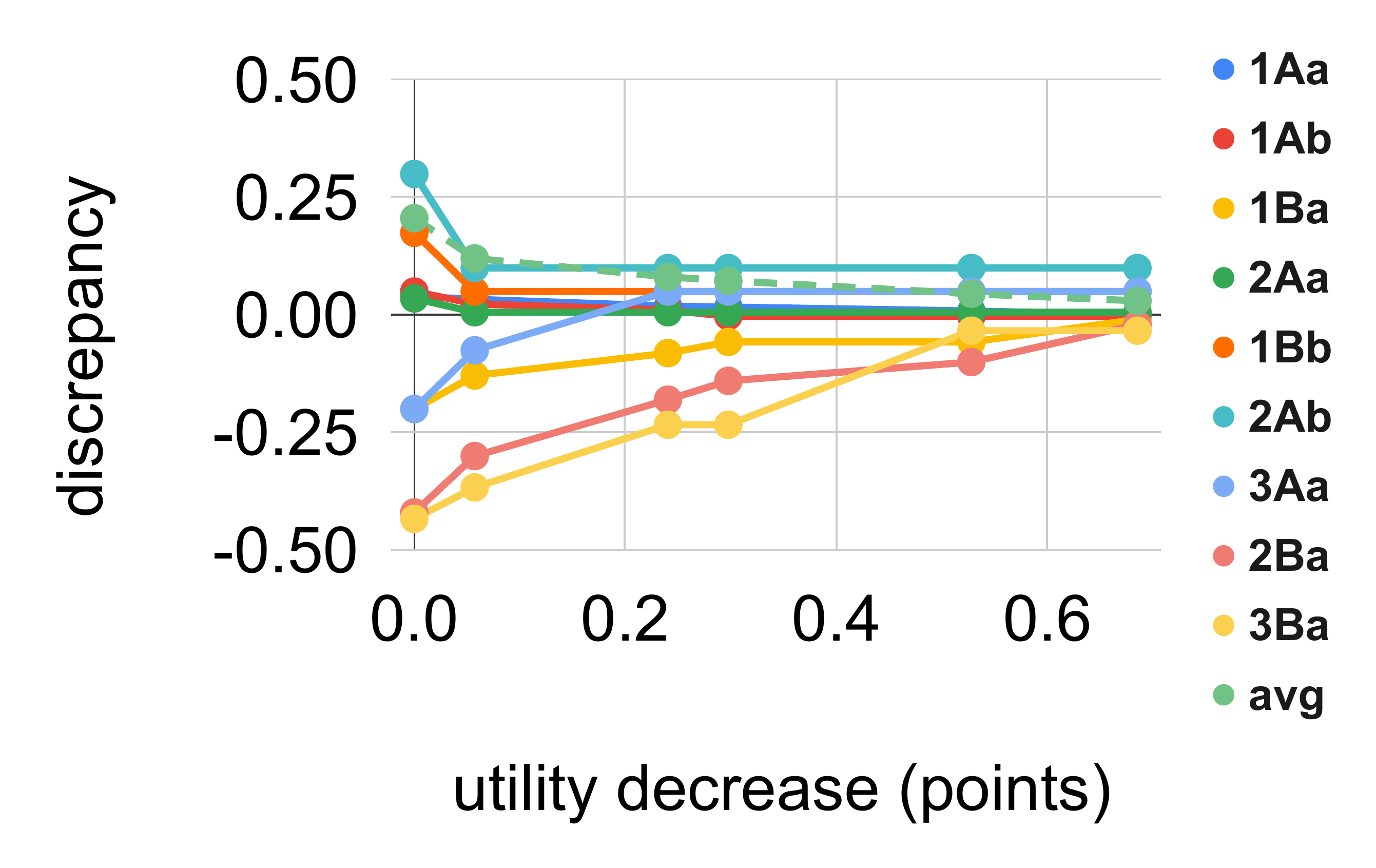}
  }
  \caption{(Best seen in color.) Per-group discrepancies in admission rates per program;
  $n$ is the number of vacancies and $\overline{u}$ the average score of those admitted to these programs.
  Programs P1-P4 are four highly selective programs (all of them are medical schools),
  while P5-P8 are four large programs having more than 350 vacancies each (two engineering schools, two business schools).
  }
  \label{fig:per-program-performance}
\end{figure*}



%% file: 07-conclusions.tex
\section{Conclusions}
\label{sec:conclusions}

In this paper, we presented a principled approach to tackle the algorithmic problem of intersectional unfairness in the context of top-k selection. We designed two different algorithms that allow to simultaneously (i) select candidates with high utility, and (ii) level up the representation of disadvantaged intersectional classes. To test our methodology, we used real data from an OECD country to simulate its 2017 university admission process. With extensive experiments, we showed that we can easily reach intersectional class parity by slightly adjusting the admission thresholds. Namely, a relatively small point decrease in the average admission score leads to almost zero unfairness. 


%% file: 08-acknowledgements.tex
\section{Acknowledgements}
\label{sec:acknowledgements}

Partially supported by the ERC Advanced Grant 788893 AMDROMA "Algorithmic and Mechanism Design Research in Online Markets" and MIUR PRIN project ALGADIMAR "Algorithms, Games, and Digital Markets".